\documentclass[aps,prl,twocolumn,longbibliography,footinbib,groupedaddress,
]{revtex4-2}

\usepackage{graphicx}% Include figure files
\usepackage{subcaption}
\usepackage{dcolumn}% Align table columns on decimal point
\usepackage{bm}% bold math
\usepackage{amsmath}
\usepackage{amssymb}
\usepackage{color}
\usepackage{xcolor}
\usepackage{hyperref}% add hypertext capabilities
\usepackage{ragged2e}
\usepackage[normalem]{ulem}

\begin{document}

\preprint{APS/123-QED}

\title{In Situ Coherence Measurements of Scattered Light in Magnetically Trapped Cold Atomic Clouds: Probe-Driven Atomic Dynamics}% 

\author{Amilson R. Fritsch$^{2}$}
\author{Hector Letellier$^{1}$}
\author{Leonardo Lima da Silva$^{2}$}
\author{Pierre Azam$^{1}$}
%\author{Michal Hemmerling$^{2}$}
\author{Vanderlei S. Bagnato$^{2}$}
\author{Robin Kaiser$^1$}
\author{Mathilde Hugbart$^1$}  \email{Contact author: mathilde.hugbart@univ-cotedazur.fr}
\affiliation{$^{1}$Universit\'e C\^ote d'Azur, CNRS, INPHYNI, France}
\affiliation{$^{2}$Instituto de Física de São Carlos, Universidade de São Paulo, São Carlos SP 13566-970, Brazil}
\date{\today}% It is always \today, today,

\begin{abstract}
The use of temporal correlations in scattered photons to probe the microscopic dynamics of ultracold quantum gases has emerged as a powerful, minimally destructive approach for \emph{in situ} analysis. Here, we demonstrate that temporal coherence spectroscopy can quantitatively characterize atomic motion in a magnetic trap, despite the perturbative effects of the probing light. By measuring the first-order correlation function $g^{(1)}(\tau)$ of light scattered by a $^{87}$Rb cloud confined in a quadrupole trap, we identify radiation-pressure-induced acceleration and heating as the origin of the apparent discrepancy between coherence spectra and temperatures inferred from time-of-flight measurements. A simple dynamical model incorporating these effects restores agreement between theory and experiment, establishing coherence spectroscopy as a reliable \emph{in situ} probe of velocity distributions in trapped atomic ensembles. Our results pave the way for time-resolved studies of nonequilibrium dynamics and thermalization processes in confined cold gases, complementing conventional destructive imaging techniques.
\end{abstract}

\maketitle

\textit{Introduction.} The use of correlations in scattered photons to probe the microscopic properties of ultracold quantum gases has emerged as a highly promising research direction. In particular, minimally destructive optical techniques offer the possibility of analyzing quantum fluids \emph{in situ}, opening the door to monitoring the temporal evolution of momentum distributions as the system undergoes relaxation. Such an approach would provide unprecedented access to the dynamics of far-from-equilibrium quantum systems, allowing the observation of processes that are otherwise difficult to capture using conventional destructive imaging methods.

Particularly relevant is the opportunity to investigate the mechanisms recently identified as distinct stages in the thermalization process~\cite{Moreno_2025}. These stages contain a wealth of information about how many-body quantum systems evolve toward equilibrium and about the specific timescales at which different physical processes begin to dominate the relaxation dynamics. Direct measurements based on photon-correlation techniques could therefore contribute significantly to a deeper understanding of nonequilibrium phenomena and the pathways through which complex quantum systems approach thermal equilibrium. Experimentally, these observables can be obtained through measurements of the first-order temporal (electric-field) correlation function $g^{(1)}(\tau)$ and the second-order (intensity) correlation function $g^{(2)}(\tau)$.

More generally, the use of temporal correlations to probe microscopic dynamics is well established in strongly scattering media through diffusing-wave spectroscopy~\cite{Pine_1988, Pine1990, Weitz1989, Fraden1990, Hebraud1997}, which provides non-invasive access to particle motion and correlated dynamics in optically opaque systems~\cite{Menon_1997, Bicout_1994}. Building on these concepts, temporal coherence measurements of light scattered by cold atomic ensembles have also already proven to be a powerful probe of light-matter interactions and atomic motion. In particular, they enable quantitative studies of the transition from single to multiple scattering in the elastic regime~\cite{Eloy_2018}, as well as the crossover from elastic to inelastic scattering at increasing probe saturation~\cite{Ortiz_2019, Lassegues_2023}. In the weak-driving regime, the temporal coherence function directly reflects Doppler-induced spectral broadening and therefore provides access to atomic velocity distributions and dynamics~\cite{Eloy_2018}. More generally, for chaotic light, first- and second-order coherence properties are connected by the Siegert relation, offering a complementary approach to characterize the field's statistical behavior~\cite{Ferreira2020}.

Extending these approaches to ultracold atomic gases is particularly appealing because coherence-based scattering measurements can probe atomic motion directly within the trapped sample. While atomic velocity distributions are most commonly inferred from time-of-flight measurements~\cite{Lett_1988}, this technique primarily provides access to the momentum distribution after expansion. Moreover, this information is obtained only in the ballistic, non-interacting limit and generally integrates over initial phase-space correlations together with interaction-driven dynamics occurring during the expansion~\cite{Gerbier_2008, Kupferschmidt_2010, Tenart_2020}. Coherence-based measurements, in contrast, can probe atomic motion \emph{in situ} and may provide access to position-dependent dynamics and velocity correlations. Such measurements could ultimately reveal spatially correlated dynamics in systems ranging from turbulent Bose-Einstein condensates~\cite{Henn_2009, Neely_2013, Moreno_2025} to dense multiple-scattering atomic clouds~\cite{Griffin_2023}.

Most coherence spectroscopy experiments with cold atoms have so far been performed either in magneto-optical traps (MOTs)~\cite{Muhammed_2016}, in optical molasses~\cite{Bali_1996, Stites_2004, Nakayama_2010, Grover_2015}, or after release from a MOT~\cite{Cipris_2026}. To our knowledge, temporal coherence measurements have not yet been reported for atoms confined in a purely magnetic trap. Such conservative trapping configurations are particularly attractive for long interrogation times and \emph{in situ} studies of ultracold gases. However, extending coherence spectroscopy to this regime introduces a central challenge: the probing light can significantly modify the atomic dynamics during the measurement. Radiation pressure and momentum diffusion can alter the velocity distribution and thereby affect the measured coherence spectrum, potentially obscuring the connection between the observed signal and the intrinsic dynamics of the atomic cloud.

In this work, we show that temporal coherence spectroscopy nevertheless remains a quantitative probe of atomic motion inside a magnetic trap when probe-induced dynamics are properly taken into account. We measure the temporal coherence of light scattered by a cloud of $^{87}$Rb atoms confined in a quadrupole magnetic trap and demonstrate that the apparent discrepancy between coherence spectra and temperatures inferred from time-of-flight expansion originates from radiation-pressure-induced acceleration and heating occurring during the measurement itself. By incorporating these effects into a simple dynamical model, we recover good agreement between theory and experiment. Our results establish coherence spectroscopy as a quantitative \emph{in situ} probe of atomic velocity distributions in trapped atomic ensembles and provide a route toward time-resolved studies of confined cold gases.

\begin{figure}\centering
    \includegraphics[width=\columnwidth]{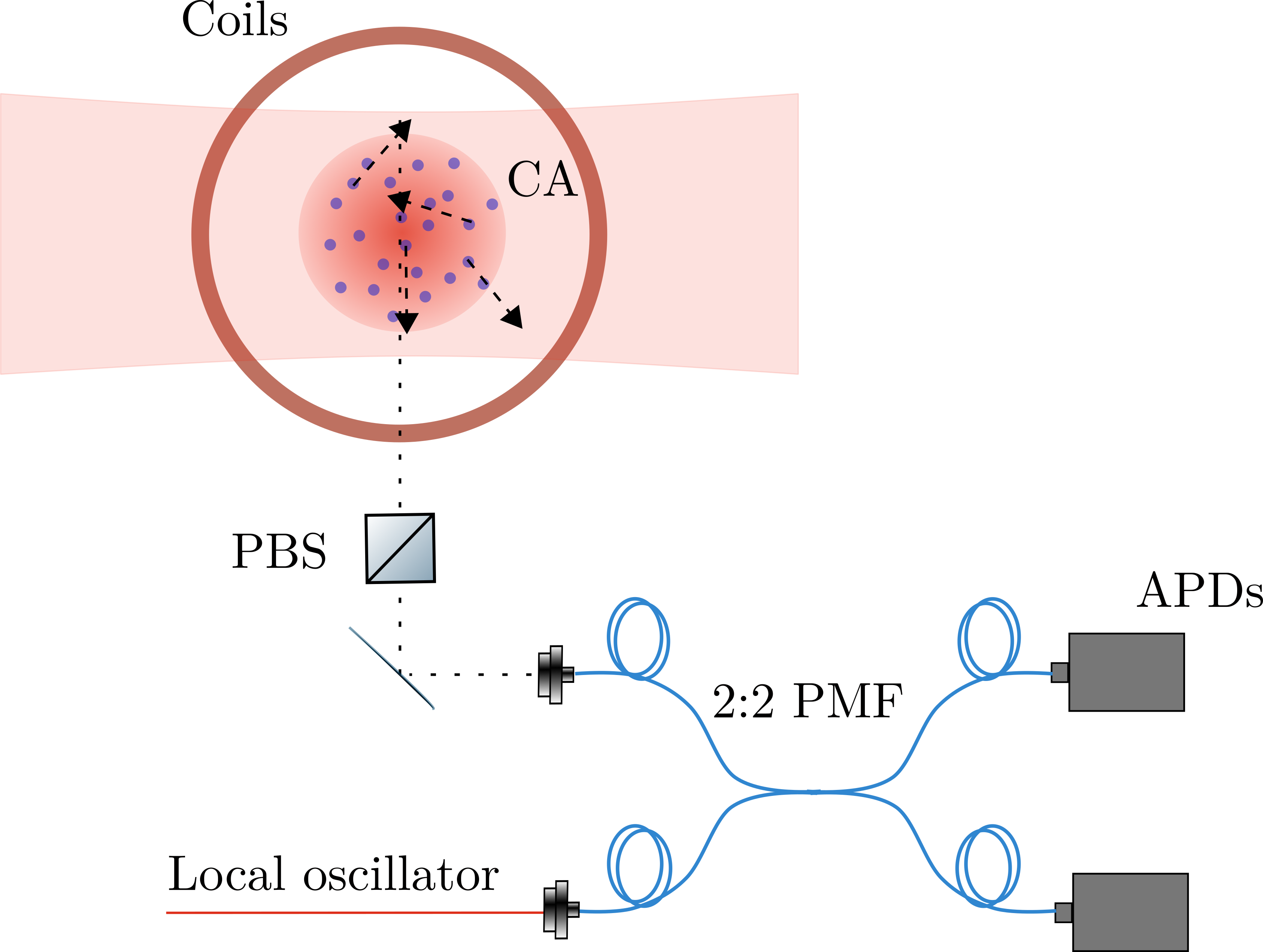}
    \caption{\justifying Experimental setup. A cold atomic cloud (CA) is confined in a quadrupole trap generated by a pair of coils located on either side of the vacuum chamber (the view in this figure is normal to the coil planes). A probe laser beam, with a waist larger than the cloud size, illuminates the atoms. Light scattered from the entire cloud (dashed arrows) is collected at $\theta = 90^\circ$, passes through a polarizing beam splitter (PBS) to select the desired polarization (see text), and is coupled into a single-mode fiber. The collected signal is then divided by a 50:50 fiber beam splitter (2×2 PMF), with each output sent to an avalanche photodiode (APD). Photon detection events are time-tagged using a time-to-digital converter and subsequently processed on a computer (not shown). The second input port of the beam splitter is used to inject a local oscillator derived from the probe, enabling measurement of the electric-field correlation function of the scattered light.}
      \label{Fig:Setup}
\end{figure}

\textit{Cold atom experimental setup.} The cold atomic ensemble consists of $^{87}$Rb prepared in the $|F = 2, m_F = 2\rangle$ state and confined in a purely quadrupole magnetic trap (see Fig. \ref{Fig:Setup}). To this end, atoms are initially loaded into a first 3D MOT and subsequently transferred to a second vacuum chamber, where they are recaptured in a second 3D MOT. After a 30\,s loading period, sub-Doppler cooling is performed by switching off the magnetic field while allowing the atoms to expand in the presence of the MOT beams. This is followed by two optical pumping pulses that transfer the population into the $|F = 2, m_F = 2\rangle$ state. All laser fields are then turned off, and the atoms are recaptured in the quadrupole trap via a 400\,ms linear magnetic field ramp. The cloud is finally allowed to thermalize for 300\,ms prior to probing.

The magnetic field gradient along the coil axis is typically set to $\Delta B_\parallel = 240$\,G/cm, with a transverse gradient given by $\Delta B_\perp = 0.44\Delta B_\parallel$. The properties of the cloud are characterized by standard absorption imaging following a time-of-flight (TOF) expansion, performed on the $F = 2 \rightarrow F' = 3$ cycling transition, and the in-situ parameters are obtained by extrapolating the TOF measurements at zero expansion time. During the quadrupole trapping stage, forced evaporation is induced by an RF knife, leading to a final temperature of $85 \pm 10\,\mu$K. The atomic density distribution is well described by a Gaussian profile, with typical in-situ root-mean-square radii $\sigma_\parallel \approx 400~\mu\mathrm{m}$ and $\sigma_\perp \approx 600~\mu\mathrm{m}$ along and perpendicular to the coil axis respectively. The total atom number is $N \simeq 10^8$, corresponding to a peak phase-space density $n_0 \lambda^3 \simeq 0.02$, where $\lambda$ denotes the wavelength of the relevant atomic transition. The system therefore remains well within the dilute regime. The collisional rate is on the order of $2\,\mathrm{s}^{-1}$, corresponding to a typical mean time between collisions of approximately $500\,\mathrm{ms}$, which is always much longer than the probe duration.

At these temperatures and cloud sizes, the characteristic motional timescale in the quadrupole trap can be estimated either from the thermal velocity, or from the typical magnetic acceleration. Both estimates yield timescales on the order of a few milliseconds. In addition, the collisional timescale is of order $500\,\mathrm{ms}$, also much longer than both the trap dynamical and probing timescales, ensuring that collisions do not play a role during the measurement. As will be discussed later, these characteristic timescales remain consistently larger than the duration of the probing sequence considered in this work.

\textit{Experimental setup for probing the coherence of scattered light.} After preparation, the atomic cloud is illuminated by a probe beam linearly polarized along the coil axis and tuned near the $\lvert F = 2 \rangle \rightarrow \lvert F' = 3 \rangle$ transition of the $D_2$ line. In the following, we assume that the dominant contribution arises from the $\lvert F = 2, m_F = 2 \rangle \rightarrow \lvert F' = 3, m_{F'} = 2 \rangle$ transition. The probe beam has a transverse size much larger than that of the atomic cloud, with a waist diameter of $2w_0 = 6.1$\,mm, ensuring a nearly homogeneous illumination.

The scattering properties are primarily determined by the optical thickness and the saturation parameter. To operate in the single-scattering regime\,\cite{Eloy_2018}, the optical thickness is kept well below unity by detuning the probe to the blue side of the atomic resonance. The optical thickness along the probe direction is given by:
\begin{equation}
b(\delta) = \frac{b_0}{1 + 4\delta^2/\Gamma^2},
\end{equation}
where $\Gamma$ is the natural linewidth and $b_0$ is the on-resonance optical thickness, extracted from TOF measurements. A blue detuning is systematically used to reduce the influence of nearby hyperfine transitions. For a typical value $b_0 \simeq 10$, a detuning $\delta \simeq 14.5\Gamma$ yields $b(\delta) = 0.012$, ensuring that the system remains mainly in the single-scattering regime.

In addition, the inhomogeneous magnetic field of the quadrupole trap induces a position-dependent Zeeman shift of the probed transition. The maximum shift, reached at the edge of the atomic cloud, is of the order of $0.7\,\Gamma$, and therefore remains smaller than the natural linewidth. This leads to a weak spatial variation of the effective detuning across the cloud, resulting in an estimated change of the optical thickness of less than $10\%$. This represents a conservative upper bound, as the magnetic field gradient predominantly affects the outer regions of the cloud, where the atomic density is lower. The system therefore remains well within the single-scattering regime under all experimental conditions considered.

The probe intensity $I$ is adjusted such that the saturation parameter remains well below unity,
\begin{equation}
s(\delta) = \frac{s_0}{1 + 4\delta^2/\Gamma^2},
\end{equation}
where $s_0 = I / I_{\mathrm{sat}}$, with $I_{\mathrm{sat}}$ the on-resonance saturation intensity. 
For linearly polarized light driving the $\lvert F = 2, m_F = 2 \rangle \rightarrow \lvert F' = 3, m_{F'} = 3 \rangle$ transition, the effective saturation intensity is $I_{\mathrm{sat}} = 1.669/(1/3)$\,mW/cm$^2$, where the factor $1/3$ accounts for the effective Clebsch–Gordan coefficient relative to the stretched-state $\sigma^\pm$ transition. Under these conditions, the scattering process remains in the linear regime and the scattered photons are predominantly elastic~\cite{Ortiz_2019}.

Scattered light is collected at an angle $\theta = 90 \pm 5^\circ$ with respect to the probe beam (see Fig.~\ref{Fig:Setup}). The scattered field contains multiple polarization components, whose relative weights depend on the underlying atomic transition and the associated selection rules (e.g. $\pi$-like contributions for $m_{F'}=2 \rightarrow m_{F}=2$ and $\sigma^\pm$ components for $m_{F'}=2 \rightarrow m_{F}=1$ channels). In the detection stage, a specific polarization component is selected by projecting the scattered field onto the linear polarization basis defined by the probe beam, using a polarizing beam splitter. The collected light is then coupled into a single-mode fiber and evenly split between two avalanche photodiodes (APDs) using a 50:50 fiber beam splitter, forming a Hanbury Brown–Twiss (HBT) configuration~\cite{HBT:1956a}. Photon arrival times are recorded by a time-to-digital converter, allowing us to construct coincidence histograms as a function of the delay $\tau$~\cite{Lassegues_2023}.

The second input port of the fiber beam splitter is used to inject a local oscillator (LO) derived from the probe beam and frequency-shifted by $f_{\mathrm{BN}} \simeq 80$\,MHz. The measured signal corresponds to the second-order correlation function $g^{(2)}_{\mathrm{BN}}(\tau)$ of the beat note between the scattered light and the LO~\cite{Ferreira2020}:
\begin{eqnarray}
g^{(2)}_{\mathrm{BN}}(\tau) &=& \frac{\langle I(t) I(t+\tau) \rangle}{\langle I(t) \rangle^2} \nonumber \\
&=& 1 + \frac{\langle I_{\mathrm{sc}}^2 \rangle}{\langle I_{\mathrm{sc}} + I_{\mathrm{LO}} \rangle^2}\left[g^{(2)}_{\mathrm{sc}}(\tau) - 1\right] \nonumber \\
&-& 2\frac{I_{\mathrm{LO}} \langle I_{\mathrm{sc}} \rangle}{\langle I_{\mathrm{sc}} + I_{\mathrm{LO}} \rangle^2}
,\left|g^{(1)}_{\mathrm{sc}}(\tau)\right| \cos(2\pi f_{\mathrm{BN}} \tau),
\end{eqnarray}
where $I_{\mathrm{sc}}$ and $I_{\mathrm{LO}}$ denote the intensities of the scattered field and the local oscillator, respectively, and $\langle \cdot \rangle$ indicates time averaging. In practice, $I(t)$ represents the photon detection events recorded by the APDs. This heterodyne HBT measurement provides simultaneous access to both the field correlation function $g^{(1)}_{\mathrm{sc}}(\tau)$ and the intensity correlation function $g^{(2)}_{\mathrm{sc}}(\tau)$ of the scattered light. These contributions are separated in Fourier space: $\tilde{g}^{(2)}_{\mathrm{sc}}$ is centered near zero frequency, while $\tilde{g}^{(1)}_{\mathrm{sc}}$ appears around the beat-note frequency $f_{\mathrm{BN}}$~\cite{Ferreira2020}.

\begin{figure}[h]
\begin{center}
\resizebox{\columnwidth}{!}{
\includegraphics{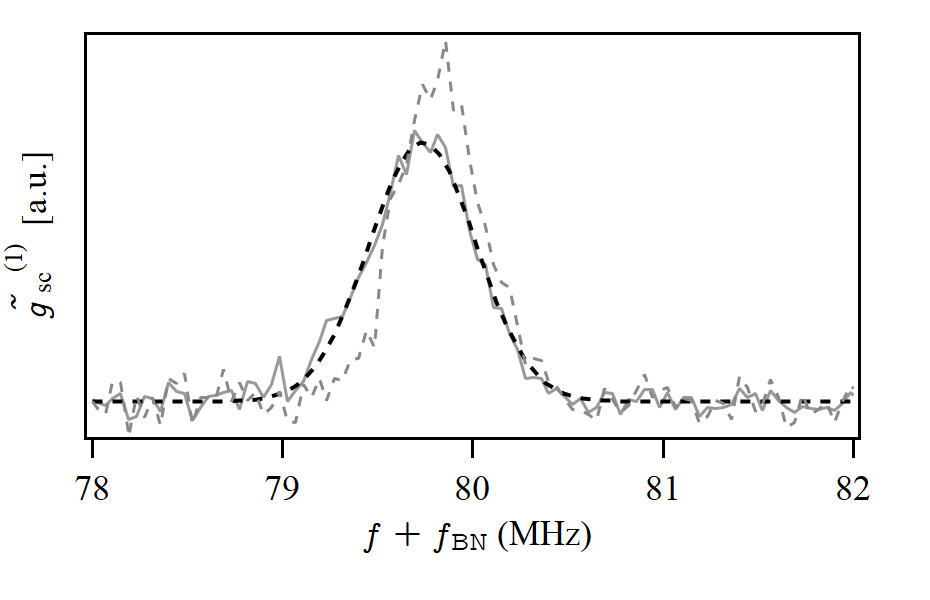}
} \caption{\justifying Grey solid curve: Fourier transform of $g^{(1)}_{\mathrm{sc}}$, centered at the beat-note frequency $f_{\mathrm{BN}} = 80\,\mathrm{MHz}$, obtained from light scattered by the atomic cloud confined in the quadrupole trap with a magnetic-field gradient of $255\,\mathrm{G/cm}$. The correlation acquisition time is $300\,\mu\mathrm{s}$ and the probe saturation parameter is $s(\delta)=0.0086$. Black dashed curve: Gaussian fit used to extract the central frequency $f_0$ and the half-width at half-maximum (HWHM) $\Delta f_1$. Grey dashed curve: Fourier transform of $g^{(2)}_{\mathrm{BN}}-1$ obtained for a shorter correlation duration of $100\,\mu\mathrm{s}$. Both experimental curves are normalized to unit area. A clear frequency shift and a reduced HWHM are observed for the shorter probe duration.} \label{Fig:FFT_fit}
\end{center}
\end{figure}

An example of the corresponding Fourier spectrum is shown in Fig.~\ref{Fig:FFT_fit} with the grey solid curve, centered around the beat-note frequency $f_{\mathrm{BN}} = 80$\,MHz and therefore associated with $\tilde{g}^{(1)}_{\mathrm{sc}}$. The data were acquired with the atomic cloud still confined in the quadrupole trap, at a magnetic-field gradient of 255\,G$/$cm. Correlation measurements were performed over a duration of 300\,$\mu$s, starting 15\,$\mu$s after the probe was switched on in order to ensure steady-state probing conditions. The spectrum is fitted with a Gaussian profile to extract both the central frequency $f_0$ and the half-width at half-maximum (HWHM) $\Delta f_1$:
\begin{equation}
    \tilde{g}^{(1)}_{\mathrm{sc}}(f) \propto e^{-\ln(2)\frac{(f-f_0)^2}{\Delta f_1^2}},    \label{Eq:Fit_g1}
\end{equation}
where the amplitude, $f_0$, and $\Delta f_1$ are treated as free fitting parameters.

The fitted parameters can be compared with theoretical expectations. First, the central frequency should coincide with the beat-note frequency. Second, in the low-saturation and single-scattering regime, the spectrum is expected to be Gaussian with a HWHM related to the cold atom temperature~\cite{Eloy_2018}:
\begin{equation}
\Delta f_T = \frac{k_\mathrm{L}}{\pi}\sqrt{\ln(2)\frac{k_\mathrm{B} T}{m}(1-\cos\theta)},
\end{equation}
where $k_\mathrm{L}$ is the probe wavenumber, $m$ is the atomic mass and $k_\mathrm{B}$ is the Boltzmann constant. Using $T = 85 \pm 10\,\mu$K, independently determined via TOF measurements, we obtain $\Delta f_T = 192 \pm 11$\,kHz. In contrast, the fitted values yield $\Delta f_1 = 332 \pm 5$\,kHz and $f_0 = 79.735 \pm 0.005$\,MHz. Both the width and the central frequency deviate from the expected values, indicating a discrepancy with the simple single-scattering thermal model.

\textit{Probe-Driven Atomic Dynamics.} Several mechanisms could account for this discrepancy. A first possibility is that the magnetic trapping field modifies the scattering properties. To test this hypothesis, we performed measurements for four different magnetic-field gradients in the range $200 < \Delta B_\parallel < 280$\,G/cm, as well as after 2 ms of time of flight, when the trapping field is switched off while the cloud parameters (size and optical thickness) remain essentially unchanged due to the short expansion time. The effect of the magnetic-field gradient can be interpreted as a position-dependent detuning of the probe across the atomic cloud, as well as the possibility of Raman processes between Zeeman sublevels (e.g. $\sigma^-$ and $\Delta m_\mathrm{F} = 1$ transition), which result in scattered photons having frequencies different from that of the incident probe light. However, no clear dependence on either the magnetic-field gradient or the presence of the trap is observed, suggesting that the origin of the discrepancy lies elsewhere in both cases.

The probe pulse duration $t_\mathrm{p}$ is chosen so as to limit the number of scattered photons per atom,
\begin{equation}
N_\mathrm{ex} = \frac{\Gamma}{2} \frac{s(\delta)}{1+s(\delta)} t_\mathrm{p},
\end{equation}
thereby minimizing radiation-pressure and heating effects. At the same time, the pulse duration must remain sufficiently long to collect enough scattered photons and ensure a satisfactory signal-to-noise ratio. As a consequence, these effects inevitably influence the measurement of $g^{(1)}$. Radiation pressure imparts a net velocity to the atomic cloud, resulting in a Doppler shift of the scattered photons, while heating increases the cloud temperature and therefore broadens the scattered-light spectrum. Both effects contribute to explain the discrepancy between the experimental observations and the theoretical predictions obtained in the absence of pushing and heating effects.

To investigate this effect, we analyzed data acquired within the same experimental run by computing the correlation functions over two different acquisition windows: the first 100\,$\mu$s and the first 300\,$\mu$s of the probe pulse. For the saturation parameters used here, ranging from $5 \times 10^{-3}$ to $8.6 \times 10^{-3}$, these integration times correspond to approximately 15 and 40 scattered photons per atom, respectively. Over the additional $\Delta t_\mathrm{p} = 200\,\mu$s interval, the atoms experience radiation-pressure forces that accelerate the cloud in the probe direction, while simultaneously undergoing heating due to photon scattering. These effects are expected to induce a shift of the central frequency of $\tilde{g^{(1)}}$ via the Doppler effect, as well as a broadening of the spectrum, both increasing with the number of exchanged photons. This behavior is indeed observed experimentally, as shown in Fig.\,\ref{Fig:FFT_fit}.

\begin{figure*}[t]
    \centering
    \begin{subfigure}[b]{0.49\textwidth}
        \includegraphics[width=\linewidth]{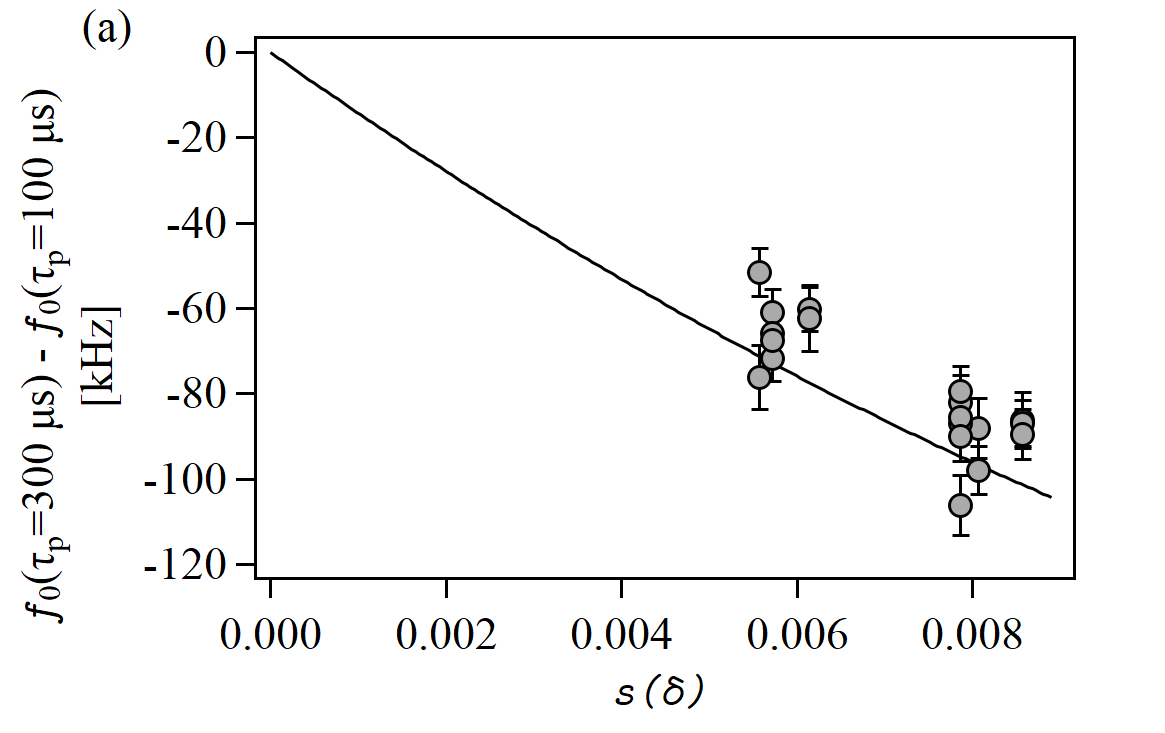}
    \end{subfigure}
    \begin{subfigure}[b]{0.49\textwidth}
        \includegraphics[width=\linewidth]{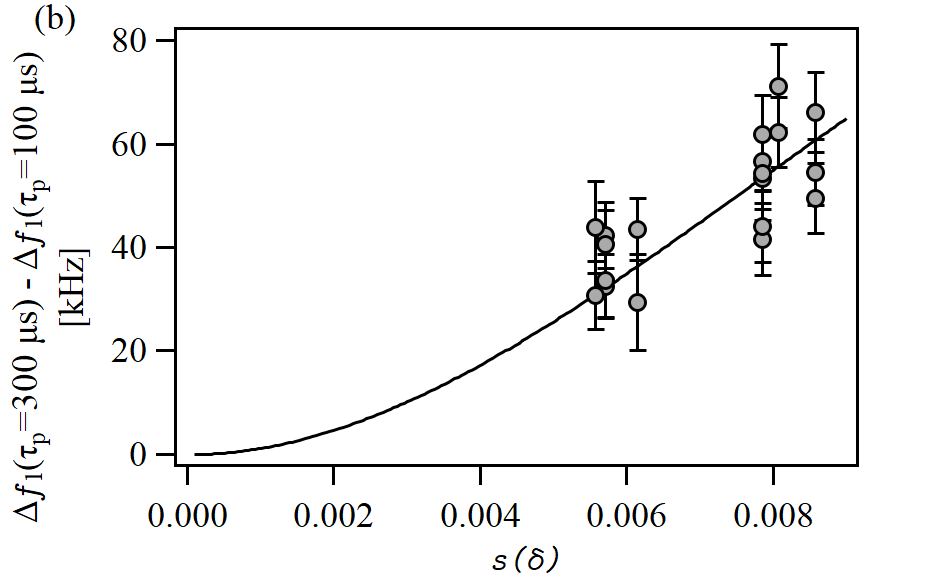}
    \end{subfigure}
    \caption{\justifying (a) Frequency shift of $\tilde{g}_{\mathrm{sc}}$ between the 100 and 300\,$\mu$s integration windows as a function of the saturation parameter. (b) Corresponding change in the linewidth of $\tilde{g}_{\mathrm{sc}}$ between the same integration times, plotted versus the saturation parameter. Symbols: experimental data. Error bars indicate $1\sigma$ uncertainties obtained from the fit of Eq.~(\ref{Eq:Fit_g1}). Solid lines: calculated frequency shift and broadening arising from radiation-pressure-induced acceleration and heating effects (see text for details).
    }
    \label{Fig:Difference}
\end{figure*}

This hypothesis is tested by evaluating the differences between the fitted parameters extracted from the two acquisition windows, namely the frequency shift $f_{0}(\tau_\mathrm{p} = 300\mu\mathrm{s}) - f_{0}(\tau_\mathrm{p} = 100\mu\mathrm{s})$ and the width change $\Delta f_1(\tau_\mathrm{p} = 300\mu\mathrm{s}) - \Delta f_1(\tau_\mathrm{p} = 100\mu\mathrm{s})$, as a function of the saturation parameter. The experimental data are shown in Fig.\,\ref{Fig:Difference}, where a clear increase of both quantities with $s(\delta)$ is observed.

Order-of-magnitude estimates are consistent with this trend. On the one hand, the net velocity induced by radiation pressure leads to a Doppler shift for the $N_\mathrm{ex}$-th scattered photon given by:
\begin{equation}
\omega_\mathrm{sc}(N_\mathrm{ex}) - \omega_\mathrm{L} = -N_\mathrm{ex} k_\mathrm{L} (1-\cos \theta) v_\mathrm{r}.
\end{equation}
where $v_\mathrm{r} = h\lambda/m = 5.8845\,\mathrm{mm/s}$ is the recoil velocity. For $s(\delta)=0.008$ and $\Delta t_\mathrm{p} = 200\,\mu\mathrm{s}$, the number of photons scattered per atom is $N_\mathrm{ex} \simeq 30$. This leads to a final Doppler shift of approximately $200\,\mathrm{kHz}$, corresponding to an average shift of about $100\,\mathrm{kHz}$ over the pulse. On the other hand, assuming a heating of $T_{\mathrm{r}}/3$ per scattering event, with $T_{\mathrm{r}} = 361.96\,\mathrm{nK}$ the recoil temperature, the cumulative temperature increase leads to a linewidth broadening on the order of a few kHz. In addition, the broadening during the pulse is also enhanced by the successive frequency shifts accumulated throughout the scattering events, which in fact constitute the dominant broadening mechanism.

Overall, this yields the correct order of magnitude, consistent with the observed frequency shift and the broadening. To obtain a more quantitative description, we employ a simplified model in which the scattered spectrum is calculated after each individual scattering event, taking into account both the Doppler shift and a temperature increase of $T_{\mathrm{r}}/3$ per event. The resulting spectra are then summed over all scattering events and fitted using the same procedure as for the experimental data, allowing us to extract the effective center frequency and linewidth.

\begin{figure}[h]
\begin{center}
\resizebox{\columnwidth}{!}{
\includegraphics{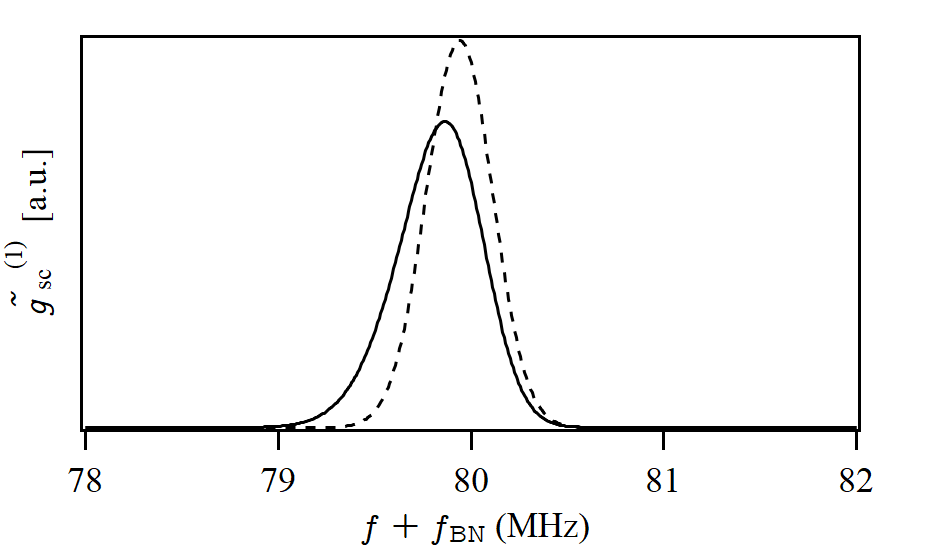}
} \caption{\justifying Dashed curve: simulated spectrum for $T = 85\,\mu\mathrm{K}$, $s = 0.008$, and $t_\mathrm{p} = 100\,\mu\mathrm{s}$. Solid curve: simulated spectrum for $t_\mathrm{p} = 300\,\mu\mathrm{s}$. Both curves are normalized to unit area.} \label{Fig:FFT_g2_BN_Simu}
\end{center}
\end{figure}

Two typical simulated spectra are shown in Fig.\,\ref{Fig:FFT_g2_BN_Simu} for $s = 0.008$. The dashed curve corresponds to a probe duration of $100\,\mu\mathrm{s}$, while the solid curve corresponds to $300\,\mu\mathrm{s}$. A clear frequency shift and linewidth broadening are observed as the probe duration increases. A slight asymmetry of the spectrum also becomes visible for longer probe times. This asymmetry originates from the fact that the total spectrum results from the sum of Gaussian contributions with different widths and amplitudes (while conserving the total area), such that the resulting lineshape is no longer strictly Gaussian. However, for the relatively small number of scattering events considered here, the effect remains weak and is therefore not observable in the experimental data given the present signal-to-noise ratio.

The fit results obtained from the simulated curves are shown in Fig.~\ref{Fig:Difference}, with the initial temperature fixed at $85\,\mu\mathrm{K}$, as independently measured via time-of-flight expansion. We find very good agreement with the experimental data, within the error bars. These results demonstrate that the properties of the atomic cloud in the magnetic trap, particularly its velocity distribution, can be reliably extracted from in-situ light-scattering measurements, provided that radiation-pressure-induced acceleration and heating effects are properly accounted for.

\textit{Perspectives.} All the previous measurements were performed using the first-order correlation function of the scattered electric field, $g^{(1)}$. The main advantage of this approach lies in its significantly better signal-to-noise ratio compared to $g^{(2)}$, which was critical in the present experiment given the particularly low duty cycle: the total experimental cycle exceeds 30\,s, while the effective acquisition window is at most 300\,$\mu$s. Even under these conditions, achieving a satisfactory signal-to-noise ratio (typically around 20) requires acquisition times of at least one full day.

Nevertheless, measurements of $g^{(2)}$ remain accessible, although with a reduced signal-to-noise ratio. An important advantage of $g^{(2)}$ is its much lower sensitivity to radiation-pressure effects, which constitute the dominant source of frequency shifts and linewidth broadening in the $g^{(1)}$ measurements. Indeed, $g^{(1)}$ is measured using a LO, such that the frequencies are determined relative to the LO frequency. In contrast, no LO is involved in the measurement of $g^{(2)}$, so the spectra remain centered around zero frequency. The residual broadening associated with heating is still present, but it remains very small, at the level of a few kHz, whereas the linewidth observed in $g^{(1)}$ reaches about $200\,\mathrm{kHz}$ due to additional broadening induced by radiation-pressure-driven acceleration.

A major strength of this approach is its capability to perform truly \emph{in situ} measurements, avoiding the need for long expansion times typically required in Bose-Einstein condensate experiments. Such a method could prove particularly valuable for systems where the dynamics evolve during expansion, for example in studies of turbulence or non-equilibrium behavior, where direct measurements inside the trap may provide more relevant information than conventional long-TOF observations. More generally, even in the absence of turbulence, this method gives direct access to the true velocity distribution, whereas time-of-flight measurements are affected by the conversion of interaction energy into kinetic energy during expansion\,\cite{Kagan_1996, Castin_1996}.

Beyond these regimes, this capability naturally motivates extending similar techniques to quantum-degenerate gases, where light scattering has recently attracted renewed interest~\cite{Lu_2023, Konstantinou_2025}. In this regime, measurements based on $g^{(2)}$ could provide complementary information beyond that contained in the scattered-light intensity alone, for instance through modifications of photon correlations induced by bosonic stimulation.

\textit{Conclusion.} In conclusion, we have demonstrated that the discrepancies between temperatures extracted from field-correlation measurements and those obtained from TOF techniques can be fully explained by radiation-pressure-induced acceleration and heating during the probe pulse. Once these effects are properly accounted for, the measured spectra are found to be in very good agreement with a simple quantitative model. Our results further confirm that the magnetic trapping field does not alter the extracted temperature values.

These findings establish light-scattering correlation spectroscopy as a reliable tool for probing the velocity distribution of atoms directly inside a magnetic trap, even in the presence of probe-induced perturbations. More generally, they demonstrate that apparent discrepancies with TOF measurements can arise from dynamical effects occurring during the probing stage rather than from intrinsic differences in the underlying atomic distribution. This work thus provides a quantitative framework for interpreting field-correlation measurements in confined geometries. While the present study focuses on first-order correlations, it also paves the way for future extensions to higher-order correlation functions.

\begin{acknowledgments}
We gratefully acknowledge Michal Hemmerling for his contribution to the initial experimental implementation of the correlation measurement setup. M.\,H. and R.\,K. acknowledge support from the ANR-FAPESP grant (ANR19-CE47-0014-01), the European Union’s Horizon 2020 programme (HALT project, Grant No. 823937), the QUANTERA ERA-NET within Horizon 2020 (PACE-IN, 8C20004, ANR19-QUAN-003-01), and the French National Research Agency (QUTISYM, ANR-23-PETQ-0002). Additional support was provided by STIC-AmSud (Ph879-17/CAPES 88887.521971/2020-00) and CAPES-COFECUB (Ph 997/23, CAPES 88887.711967/2022-00) together with H.\,L. and P.\,A.. M.\,H. also acknowledges support from the UCA J.E.D.I. Investments programme (ANR-15-IDEX-01), while R.\,K. acknowledges funding from the European Research Council under the Advanced Grant ANDLICA (No. 832219) and from the ANR (LiLoA, ANR-23-CE30-0035). V.\,S.\,B.,  A.\,R.\,F. and L.\,L.\,S. acknowledge support from the FAPESP grant (2013/07276-1) and A.R.F also acknowledge support from FAPESP grant 2024/21658-9.
\end{acknowledgments}

\bibliography{Biblio}{}% Produces the bibliography via BibTeX.

\end{document}